\documentclass[iop,apjl]{emulateapj}

\hbadness = 10000 \usepackage{natbib}
\bibpunct{(}{)}{;}{a}{}{,} 
\usepackage[usenames,dvipsnames]{xcolor}
\usepackage[breaklinks,colorlinks,linkcolor=Maroon,urlcolor=Blue,citecolor=Blue]{hyperref}
\usepackage{amsmath}
\usepackage{amssymb}
\usepackage[utf8]{inputenc}
\usepackage[T1]{fontenc}
\usepackage[english]{babel}
\usepackage{graphicx}
\usepackage{xspace}
\usepackage{enumerate}
\usepackage[all]{hypcap} 
\usepackage{soul}
\setulcolor{Red}
\usepackage[colorinlistoftodos,textsize=tiny]{todonotes} \usepackage[capitalise,nameinlink,noabbrev]{cleveref}

\makeatletter
\usepackage{etoolbox}
\patchcmd\H@refstepcounter{\protected@edef}{\protected@xdef}{}{}
\makeatother

\newcommand{\alphat}{\ensuremath{\alpha_\mathrm{t}}\xspace}
\newcommand{\cs}{\ensuremath{c_\mathrm{s}}\xspace}
\newcommand{\af}{\ensuremath{a_\mathrm{f}}\xspace}
\newcommand{\vf}{\ensuremath{v_\mathrm{f}}\xspace}
\newcommand{\rf}{\ensuremath{r_\mathrm{f}}\xspace}
\newcommand{\ad}{\ensuremath{a_\mathrm{d}}\xspace}

\newcommand{\tdisk}{\ensuremath{t_\mathrm{disk}}\xspace}

\newcommand{\eps}{\ensuremath{\epsilon}\xspace}

\newcommand{\Ok}{\ensuremath{\Omega_\mathrm{k}}\xspace}

\newcommand{\rhos}{\ensuremath{\rho_\mathrm{s}}\xspace}
\newcommand{\Sigd}{\ensuremath{\Sigma_\mathrm{d}}\xspace}
\newcommand{\Sigg}{\ensuremath{\Sigma_\mathrm{g}}\xspace}

\newcommand{\St}{\ensuremath{\mathrm{St}}\xspace}
\newcommand{\Sc}{\ensuremath{\mathrm{Sc}}\xspace}

\newcommand{\Vk}{\ensuremath{V_\mathrm{k}}\xspace}

\slugcomment{Received 2015 September 10; accepted 2015 October 12}
\shorttitle{\textsc{Dust Evolution Can Produce Scattered Light Gaps}}
\shortauthors{\textsc{Birnstiel et al.}}

\begin{document}
\title{Dust Evolution Can Produce Scattered Light Gaps in Protoplanetary Disks}
\author{Tilman Birnstiel\altaffilmark{1,3}}
\author{Sean M. Andrews\altaffilmark{1}}
\author{Paola Pinilla\altaffilmark{2}}
\author{Mihkel Kama\altaffilmark{2}}

\affil{\altaffilmark{1}Harvard-Smithsonian Center for Astrophysics, 60 Garden Street, Cambridge, MA 02138, USA\\
\altaffilmark{2}Leiden Observatory, P.O. Box 9513, 2300 RA, Leiden, The Netherlands}

\altaffiltext{3}{present address: Max Planck Institute for Astronomy, K\"onigstuhl 17, 69117 Heidelberg, Germany}

\email{tbirnstiel@cfa.harvard.edu}
\email{sandrews@cfa.harvard.edu}
\email{pinilla@strw.leidenuniv.nl}
\email{mkama@strw.leidenuniv.nl}

\begin{abstract}
Recent imaging of protoplanetary disks with high resolution and contrast have revealed a striking variety of substructure. Of particular interest are cases where near-infrared scattered light images show evidence for low-intensity annular ``gaps.'' The origins of such structures are still uncertain, but the interaction of the gas disk with planets is a common interpretation. We study the impact that the evolution of the solid material can have on the observable properties of disks in a simple scenario without any gravitational or hydrodynamical disturbances to the gas disk structure. Even with a smooth and continuous gas density profile, we find that the scattered light emission produced by small dust grains can exhibit ring-like depressions similar to those presented in recent observations. The physical mechanisms responsible for these features rely on the inefficient fragmentation of dust particles. The occurrence and position of the proposed ``gap'' features depend most strongly on the dust-to-gas ratio, the fragmentation threshold velocity, the strength of the turbulence, and the age of the disk, and should be generic (at some radius) for typically adopted disk parameters. The same physical processes can affect the thermal emission at optically thin wavelengths ($\sim$1\,mm), although the behavior can be more complex; unlike for disk--planet interactions, a ``gap'' should not be present at these longer wavelengths. 
\end{abstract}

\keywords{accretion, accretion disks --- circumstellar matter ---  protoplanetary disks}

\section{Introduction}\label{sec:intro}

Advances in high-contrast imaging techniques and equipment have put the direct detection of a forming planet embedded in a circumstellar disk within reach \citep[see][]{Kraus:2012p16079,Quanz:2013p20189,Quanz:2015p24515,Biller:2014p24400,Currie:2014p24520,Reggiani:2014p24431}. Along the path toward that goal, observing campaigns that exploit these advances have revealed that a variety of disk substructures can be seen in scattered light images, including spiral patterns \citep[e.g.,][]{Muto:2012p20271}, and gaps \citep[e.g.,][]{Debes:2013p23285}. Given the ubiquity of planetary systems found to be orbiting main-sequence stars \citep[e.g.,][]{Batalha:2013p24469}, it is reasonable to presume that these features are generated by various interactions between young planetary systems and the disks they form in \citep[see the recent review by][]{Baruteau:2014p23060}.

But there are several alternative explanations for the types of substructure observed in these disks that do not necessarily rely on dynamical interactions with planets. Some are geometric, particularly in cases where the radial dependence of the scattering surface is affected by shadowing 
\citep[e.g.,][]{Quillen:2006p24478,Siebenmorgen:2012p23483,Marino:2015p24512}. Others invoke hydrodynamic processes, including vortices at dead zone edges \citep[e.g.,][]{Varniere:2006p24475,Regaly:2012p18108} or a photoelectric instability \citep[e.g.,][]{Klahr:2005p20766,Lyra:2013p21037}. Spatial variations in the properties of the scatterers --- dust grains embedded in the gas disk --- are also frequently noted as a potential explanation. However, this latter possibility has not yet been explored in much detail for this specific context.

To explore this last avenue further, this Letter discusses the observational impact that ``normal'' dust evolution has on scattered light observations of disks that are neither hosting a planet nor subject to any of the hydrodynamic instabilities mentioned above. The goal is to determine the observable substructure that can be caused by particle size variations due to the transport and growth of solid particles. In \Cref{sec:globaldistributions}, we review the key concepts and assumptions on which this study rests. In \Cref{sec:discussion}, we investigate the observable consequences and discuss some constraints on physical parameters. We provide a repository\footnote{See also, \url{http://birnstiel.github.io/Birnstiel2015_scripts/}} \citep{10.5281/zenodo.32488} with all the data and code to reproduce our results.

\section{Global distributions of particle sizes}\label{sec:globaldistributions}

First we briefly review the basic concepts and assumptions in the dust evolution framework of \citet{Birnstiel:2012p17135}. We assume that a collision between solid particles leads to perfect sticking if the impact velocity is below a threshold velocity \vf, and to fragmentation/erosion at higher relative velocities. The size-dependent particle motions that induce these collisions and drive the transport and evolution of the disk solids are determined by three key processes: (1) turbulent mixing, (2) coupling to the gas flows driven by viscous accretion and spreading, and (3) radial drift toward the gas pressure maximum \citep{Weidenschilling:1977p865}. For the mechanisms considered here, the impact velocities increase for larger particle sizes. A more complete summary is available in the review by \citet{Testi:2014p23004}.

Solid particles influenced by these processes will grow into larger aggregates until either they collide at high velocities and fragment or they drift toward the star faster than the collision timescale. These two obstacles to further growth are known as the \emph{fragmentation barrier} and the \emph{drift barrier}, respectively. The limiting particle sizes that correspond to these barriers, \af and \ad, respectively, are 
\begin{align}
\af &\simeq \frac{2}{3\pi} \, \frac{\Sigg}{\rhos \,\, \alphat} \, \frac{\vf^2}{\cs^2}\label{eq:a_frag}\\
\ad &\simeq \frac{2}{\pi} \, \frac{\Sigd}{\rhos \left|\gamma\right|}\frac{\Vk^2}{\cs^2} \label{eq:a_drift}
\end{align}
\citep[cf.,][]{Birnstiel:2012p17135}.
Here $\Sigma_\mathrm{d}$ and $\Sigma_\mathrm{g}$ are the dust and gas surface densities, \cs the sound speed, \alphat the turbulent viscosity parameter \citep{Shakura:1973p4854}, \rhos the bulk density of the particles, \Vk the Keplerian velocity, and $\gamma = \partial\ln P/\partial\ln r$ the pressure gradient. 

For a typical disk structure, there is a limited region of parameter-space where fragmentation operates efficiently. Fragmentation velocities estimated from laboratory experiments are $\sim$few $\times$1--10\,m~s$^{-1}$, depending on the composition and internal structure of the particles \citep{Blum:2000p8099,Wada:2009p8776,Guttler:2010p9745,Gundlach:2015p24161}. Such relative velocities are prevalent in the inner part of the disk. Radial drift itself does not facilitate impact velocities comparable to \vf for long times; eventually such motions diminish $\Sigma_\mathrm{d}$, and thereby \ad, to the point that the relative velocities are below \vf. Turbulent mixing can in principle enhance fragmentation, although it too has restricted impact. The maximum relative velocity induced by turbulence is 
\begin{equation}
\Delta v_\mathrm{max} \approx \cs\,\sqrt{\frac{3}{2}\alphat}
\label{eq:v_max}
\end{equation}
\citep{Volk:1980p19798,Ormel:2007p801}. Since $\Delta v_\mathrm{max}$ decreases with distance from the star ($r$), it will drop below \vf unless \cs and/or \alphat remain high. This implies that drift makes it difficult to sustain fragmentation at all disk radii, unless the turbulence is strong \citep{Birnstiel:2009p7135} or \vf is much lower than expected. 

\begin{figure}[t]
\includegraphics[width=\hsize]{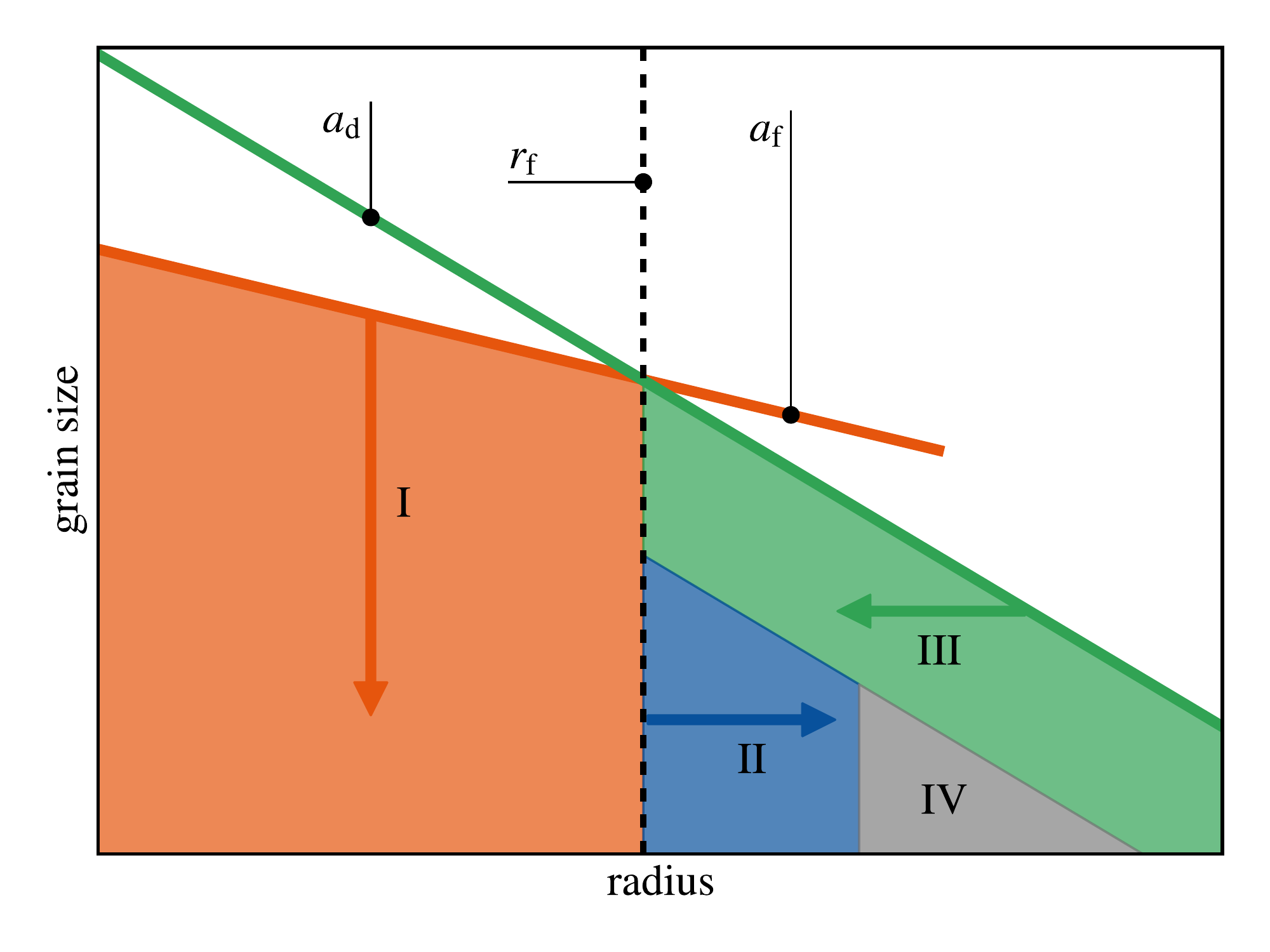}
\figcaption{Schematic description of the global distribution of grain sizes as a function of disk radius. Particles grow toward \af (orange line) or \ad (green line), depending on the local physical conditions. In the inner disk (region I), fragmentation of the largest particles replenishes the population of small dust (denoted by a downward arrow). In the outer disk, fragmentation is not effective and the grain size distribution becomes strongly top-heavy (with a much reduced population of small particles). Radial diffusion transports small grains from the inner regions outward (outward arrow, region II) and also radially smears out the top-heavy size distribution in the outer disk (inward arrow, region III). However, this leaves a radial band (region IV) that lacks small particles.
\label{fig:scematic}}
\end{figure}

Based on these considerations, there are some basic global trends in the distribution of particle sizes as a function of radius that should be expected from the growth and evolution of disk solids. These are highlighted in \Cref{fig:scematic}, which breaks them down into regions in the radius--grain size parameter-space, and can be summarized as follows. 
\begin{enumerate}[I)]
  \item In the hot, inner region ({\it orange} in \cref{fig:scematic}), particles grow up to \af and then fragment. The particle size distribution reaches an equilibrium between coagulation and fragmentation, which is slightly top-heavy with most of the mass in particles near \af, but still includes copious amounts of small grains. 
  \item While fragmentation is only effective interior to a radius \rf (region I), some small fragments that are continuously supplied there will be radially mixed outward by turbulent diffusion ({\it blue} in \cref{fig:scematic}). 
  \item In the outer regions ({\it green} in \cref{fig:scematic}), growth is limited by (inward) drift. There, the particle size distribution is much more top-heavy, because growth and drift conspire to concentrate particle sizes around \ad \citep[][see their Fig.~6]{Birnstiel:2012p17135}. Nevertheless, these distributions will have a finite width due to radial diffusion. 
  \item Consequently, there is a region where a population of small particles is not replenished: fragmentation is inefficient, turbulent diffusion does not extend there (from region II), and the growth/drift timescales are short. Moreover, because most of the solid mass in that radial zone is concentrated at larger sizes (region III), any small particles that might be present would be quickly swept up by the population of larger, drifting solids. This distinctive region of parameter-space is colored gray in \Cref{fig:scematic}, and will be the focus here.
\end{enumerate}

The general behavior described above, including the region depleted of small particles noted in IV, have already appeared in various numerical simulations of particle evolution in protoplanetary disks \citep[e.g.,][]{Brauer:2008p215,Laibe:2008p19156,Birnstiel:2010p9709,Birnstiel:2012p17135,Pinilla:2012p18741}. Similar behavior is also predicted in models that consider porosity evolution in the coagulation process \citep[e.g.,][]{Okuzumi:2012p18448,Krijt:2015p24077}. In the following sections, we discuss the criteria that determine the boundaries of these four general regions of parameter-space, and develop a semi-analytical treatment to approximate these complex simulations.

\subsection{Efficient Fragmentation (Region I)}

We make the assumption that solid particles will only fragment efficiently in the inner region of a disk, where impact velocities are high. This should generally be the case, since those velocities scale with either \cs (in the case of Brownian motion or turbulent mixing) or \Vk (for vertically settling and radial or azimuthal drift), both of which decrease with $r$. We define \rf as a critical radius in the problem: fragmentation is suppressed beyond it. As highlighted above, there are two possible reasons for that suppression. First, the maximum collision speeds between particles could be lower than required for fragmentation; i.e., $\Delta v_\mathrm{max} < \vf$. Second, and more important, the particles could achieve high enough drift speeds that they move inwards before they can reach the sizes (and thereby relative velocities) at which fragmentation is effective. This latter case is the ``drift limit,'' where $\ad < \af$; it can be reformulated using Eq.~\cref{eq:a_frag,eq:a_drift} as
\begin{equation}
\eps \, \alphat < \frac{|\gamma|}{3} \left(\frac{\vf}{\Vk}\right)^2,
\label{eq:rf_criterion}
\end{equation}
where $\eps = \Sigma_\mathrm{d}/\Sigma_\mathrm{g}$ is the dust-to-gas mass ratio. $\gamma$, \eps, and \alphat are not expected to be uniform throughout the disk, but by assuming they are, we can roughly estimate the maximum radius that fulfills this condition,
\begin{equation}
\frac{\rf}{100\,\mathrm{AU}} \sim \frac{M_\star}{M_\odot} \, \frac{\alphat}{10^{-3}} \, \frac{\eps}{0.01} \, \left(\frac{|\gamma|}{2.75}\right)^{-1}\,\left(\frac{\vf}{10\,{\rm m~s}^{-1}}\right)^{-2}.
\label{eq:rf_estimate}
\end{equation}
Radial drift can significantly reduce $\eps$ in the outer disk, thereby pushing $\rf$ well below 100~AU. Lower limits on \rf are likely given by sintering \citep[e.g.,][]{Sirono:2011p21039} or by a reduced \vf inward of the water evaporation front where particles loose their ice mantles.

The particle size distribution inside \rf (region I) will quickly reach a coagulation-fragmentation equilibrium, for which analytical fits have been derived by \citet{Birnstiel:2011p13845}. Here we approximate these as
\begin{equation}
\Sigd(r, a) =
\begin{cases}
C(r)\cdot a^{1/2}
	& \mathrm{for}\, a \leq a_\mathrm{BT}\\
\Sigd(r,a_\mathrm{BT}) \cdot \left(\frac{a}{a_\mathrm{BT}}\right)^{-5/8}
	& \mathrm{for}\, a_\mathrm{BT} < a < \af
\end{cases},
\label{eq:frag_distri}
\end{equation}
where $a_\mathrm{BT}$ (itself a function of radius) is the transition from Brownian motion to turbulent relative velocities (cf., Eq.~37 in \citealt{Birnstiel:2011p13845}), and $C(r)$ is a normalization constant such that $\Sigd(r) = \int \Sigd(r, a)\,da$.

\subsection{Radial Diffusion of Fragments (Region II)}

Even though fragmentation is inefficient outside \rf, larger radii can still be supplied with small particles (from region I) due to outward radial diffusion. If we consider such particles to be passive tracers of the gas, we can apply the analytical treatment of \citet{Pavlyuchenkov:2007p802} to predict their concentration outside \rf. In this scenario, the surface density of a given particle size can be approximated in steady state with a power-law distribution,
\begin{equation}
\frac{\Sigd(r, a)}{\Sigg(r)} \propto \left(\frac{r}{\rf}\right)^{-\frac{3}{2}\,\Sc} \,\text{for}\quad \rf<r<r_0,
\label{eq:diffusion}
\end{equation}
where \Sc is the Schmidt number, the ratio of the particle diffusivity to the gas diffusivity. Here we assume \Sc = 1 for all particle sizes. $r_0$ is the radius at which the particle size $a$ equals the drift limit \ad.

\subsection{Drift-dominated Evolution (Region III)}

Outside of \rf, particles grow toward a limiting size, \ad, defined by equality between the drift and growth timescales. Most of the dust mass then resides close to \ad (see \Cref{fig:scematic}). However, radial diffusion effectively broadens the particle size distribution: even if particle growth were to produce a hypothetical monodisperse size distribution like $\delta(a-\ad)$ at a given $r$, radial diffusion will move particles from larger $r$ inwards to that radius. Those particles will be smaller due to the radial dependence of \ad, and therefore the size distribution will necessarily broaden in the presence of turbulent diffusion. This inward mixing is balanced by sweep-up collisions with drifting particles of size \ad. To approximate these effects, we will parametrize the radial dependence of the surface density for a given grain size as a power law with index $p_\mathrm{d}$. This index can be estimated from balancing drift, diffusion, and sweep-up collisions (Birnstiel et al. 2016, in preparation); it generally depends on various physical parameters of the disk, but in the following, we take it to be $p_\mathrm{d}=2.5$ (the results of interest here are rather insensitive to $p_\mathrm{d}$).

The same principles also apply to the outward diffusion of the drifting particles. In this case, radial drift is opposing that outward diffusion. Similar to the case in region II, we can find a steady state solution for this scenario, but since radial drift velocities are much higher than the diffusion velocities, this outward motion of drifting particles is highly ineffective. Solving for the stationary solution analogous to \citet[][ see also \citealp{Jacquet:2012p19670}]{Pavlyuchenkov:2007p802} then defines the second part of our approximation, where
\begin{equation}
\frac{\Sigd(r,a)}{\Sigd(r_0,a)} =
\begin{cases}
\left(\frac{r}{r_0}\right)^{p_\mathrm{d}}													& \mathrm{for}\,r \leq r_0 \\
\frac{\Sigg(r)}{\Sigg(r_0)}\cdot\exp\left(\int_{r_0}^r \frac{\St\,\gamma}{\alphat}dr'\right)	& \mathrm{for}\,r>r_0\\
\end{cases}.
\label{eq:outer_diffusion}
\end{equation}
Here $\St = a\,\rhos\,\pi/2\Sigg$ denotes the Stokes number at the disk mid-plane.

\begin{figure}[t]
\includegraphics[width=\hsize]{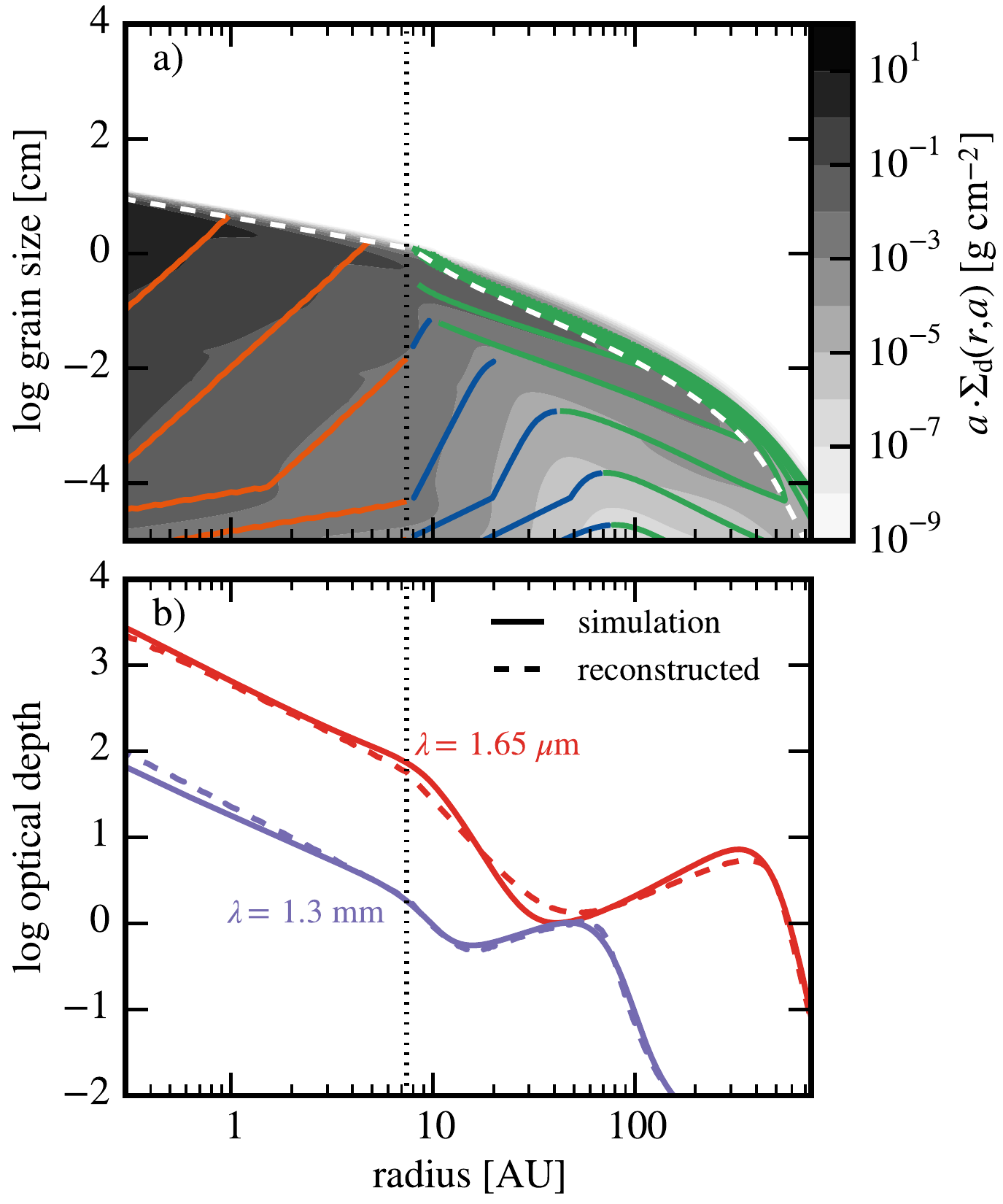}
\figcaption{Comparison of the simulated and semi-analytic size distribution as a function of disk radius for an illustrative example. Panel (a): simulation result after 1\,Myr (gray scale). The dashed and dotted lines represent \af and \rf, respectively. The approximate, reconstructed size distribution is overlaid as contour levels with colors matching the regions in \cref{fig:scematic}. Panel (b): vertically integrated optical depth (absorption + scattering) profile at wavelengths of 1.65~$\mu$m and 1.3~mm for the simulated distribution (solid) and its reconstruction (dashed).
\label{fig:comparison}}
\end{figure}

\subsection{Synthesis: Region IV}
Given a set of disk parameters, \cref{eq:rf_criterion,eq:rf_estimate,eq:frag_distri,eq:diffusion,eq:outer_diffusion} can be used to reconstruct the particle size distribution as a function of radius. The results for an illustrative example\footnote{The simulation setup follows \citet{Birnstiel:2012p17135} but uses the following parameters: $M_\star=0.7\,M_\odot$, $M_\mathrm{disk} = 0.1\,M_\star$, $r_\mathrm{c} = 200$~AU, $\vf = 10$~m~s$^{-1}$, and $\alphat=10^{-3}$.} are shown in \Cref{fig:comparison}. Panel (a) shows the full simulation after 1\,Myr (using the \citealt{Birnstiel:2010p9709} code) in gray scale. The framework explained above was used to reconstruct the size distribution, given \Sigg and \Sigd from the simulation. The results are overlaid as contour lines at the same levels; colors correspond to the respective regions in \cref{fig:scematic}. The key features are reproduced by this approximation: copious amounts of small grains reside inside \rf and extend to larger radii, corresponding to regions I and II, respectively.
The densities for small grains in region II decrease with $r$ (reflecting the limits of turbulent diffusion), while in region III they increase with $r$ (due to longer growth timescales and smaller \ad). This gives rise to a small-grain density {\it minimum} at intermediate radii (region IV). The radial segregation of particle sizes (\ad decreases with $r$) means that the location of this minimum decreases with $a$.

Panel (b) aims to connect this density minimum feature for small particles to a more observationally oriented diagnostic. It shows the radial distribution of (vertically integrated) optical depths for the simulated and reconstructed (approximated) models at wavelengths of $1.65\,\mu$m and 1.3\,mm, where spatially resolved data are relatively common for protoplanetary disks. The reconstruction generally approximates the more detailed simulation remarkably well in terms of this key diagnostic.

\section{Discussion}\label{sec:discussion}

We have illustrated that the normal evolution of the solid particles embedded in a gas-rich disk should impart some pronounced substructure -- a radial zone where small particles are preferentially depleted -- that can in principle be measured and characterized. Here we comment on the feasibility of that characterization, as well as on the region of parameter-space where these features are expected to occur.

\begin{figure}[th]
\includegraphics[width=\hsize]{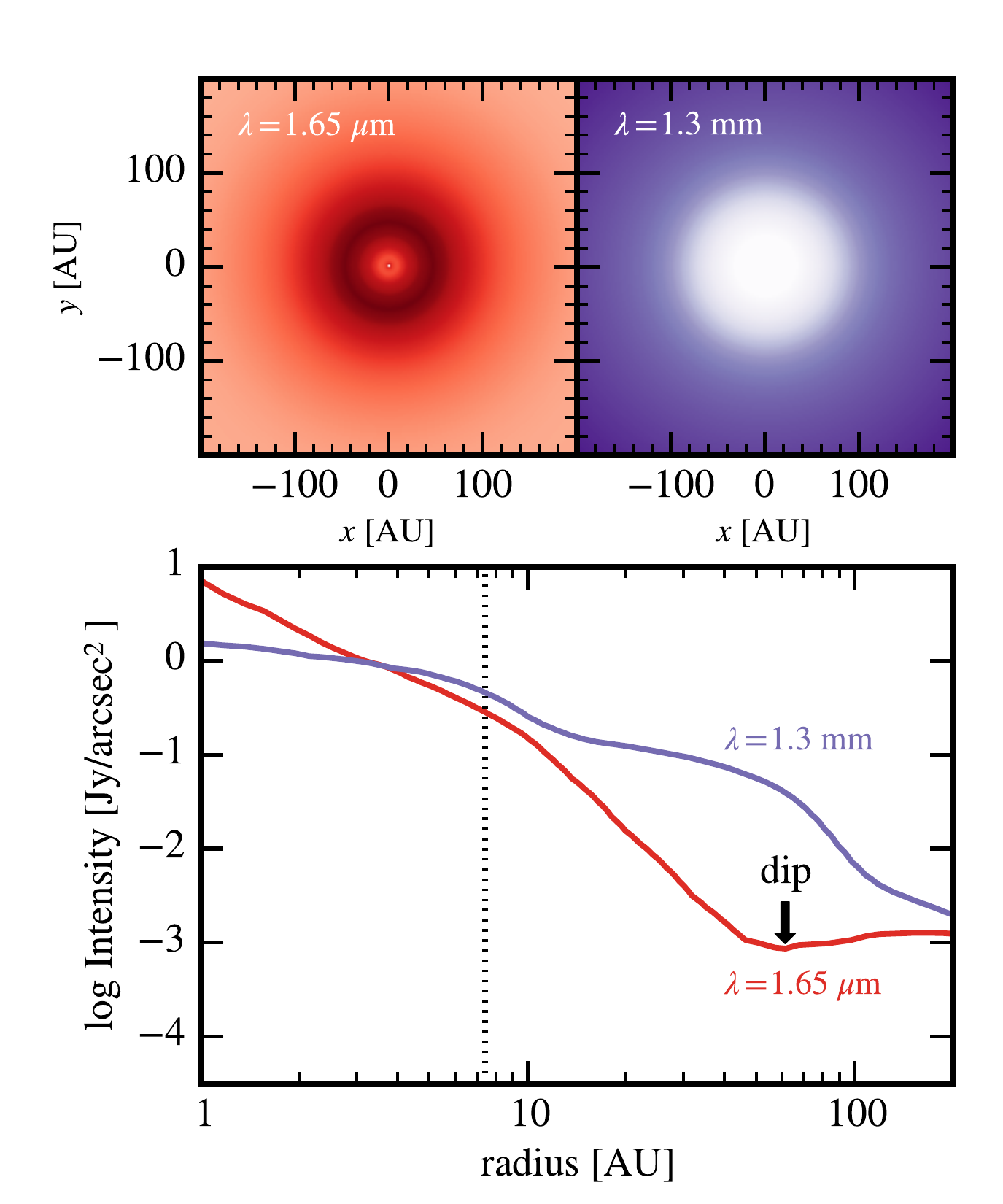}
\figcaption{({\it top}) Synthetic images at 1.65~$\mu$m (left) and 1.3\,mm (right) for the simulation shown in \cref{fig:comparison} (both with a logarithmic stretch to the color-scale). The scattered light image at 1.65\,$\mu$m has been scaled by $r^2$, a common practice to remove the effects of a diluted stellar radiation field. ({\it bottom}) The corresponding azimuthally averaged radial intensity profiles. The dotted line marks \rf.
\label{fig:profiles}}
\end{figure}

\subsection{Observable Signatures}\label{sec:discussion:observations}

To evaluate the observable consequences of this rather general feature in current models of dust evolution, we used the simulation outputs to calculate images and radial intensity profiles using the {\tt RADMC-3D} radiative transfer code \citep{Dullemond:2012p24314}. To do that, we assumed the gas density is in vertical hydrostatic equilibrium, with the dust density of each particle size distributed according to a mixing-settling equilibrium \citep[e.g.,][]{Schrapler:2004p2394}. The vertical density structure was iterated based on the output thermal structure from {\tt RADMC-3D} following the approach of \citet{Kama:2009p23727}, allowing a maximum dust density deviation of 10\%\ in each iteration to avoid numerical instabilities. We also employed the partial diffusion approximation and modified random walk procedures advocated by \citet{Min:2009p23566}. We adopted the dust opacities prescription from \citet{Ricci:2010p9423}, and assumed isotropic scattering for simplicity.

The raytraced images and (azimuthally averaged) intensity profiles at wavelengths of 1.65\,$\mu$m ($H$-band) and 1.3\,mm for the simulation of \Cref{fig:comparison} are shown in \Cref{fig:profiles}. Notably, the significant reduction in the optical depth profile produces a depression in the scattered light intensity, starting roughly at \rf (in this case $\sim$7.4\,AU) and reaching a minimum at about 50~AU. The tendency of this depression is to move inward with time because radial drift is depleting \Sigd, thus reducing \eps, which in turn decreases \rf according to \cref{eq:rf_estimate}. The depression noted in the scattered light intensity resembles the recently observed ``gap'' features seen in the disks around TW Hya \citep{Debes:2013p23285,Akiyama:2015p24232} and HD~169142 \citep{Quanz:2013p20189}. Such a feature is often linked to a planetary mass companion, which can induce a similar pattern in the scattered light morphology \citep[e.g.,][]{deJuanOvelar:2013p23197}. However, in this paper the dip is purely caused by opacity effects: the total dust surface density is continuous.

In principle, one could distinguish between these two potential origins for such a feature with resolved data at an optically thin wavelength like 1.3\,mm. There, the ``gap'' seen in scattered light will be absent if it is caused solely by dust evolution. Still, the intensity profile at 1.3\,mm may also exhibit some more complex behavior outside of \rf, where the lack of fragmentation leads to a reduced surface density of mm-sized particles. A second drop in mm-wave intensity at even larger radii may be caused by a lack of large particles in the outer disk. The locations and shapes of such features depends strongly on the choice of disk parameters: special care must be taken in their interpretation (the example presented here is not necessarily representative behavior at 1.3\,mm).

\subsection{Range of Applicability}\label{sec:discussion:caveats}

\begin{figure}[t]
\includegraphics[width=\hsize]{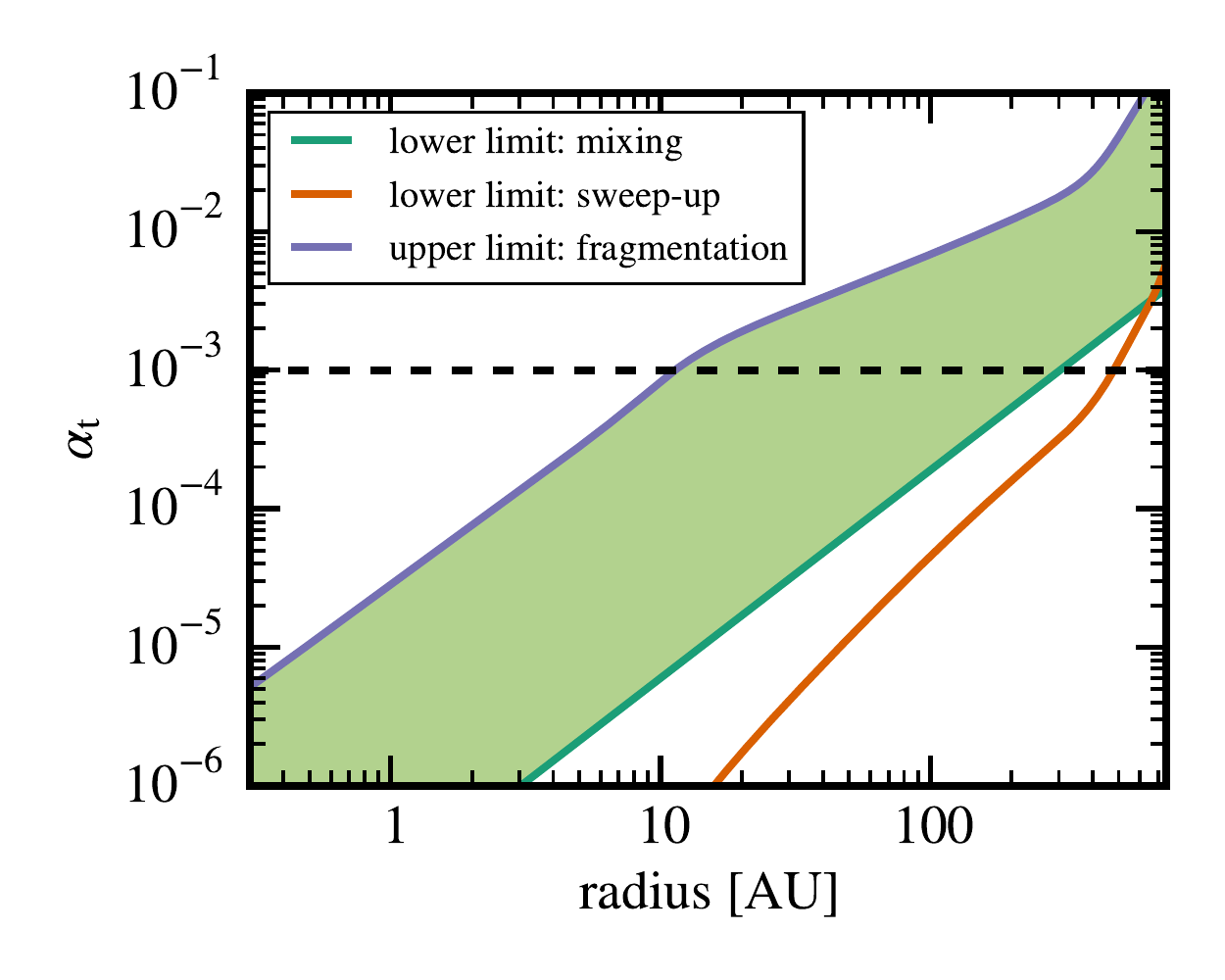}
\figcaption{Constraints on \alphat for the illustrative model discussed here to produce a ``gap'' feature in the scattered light emission. The limits on \alphat are imposed constraints on the vertical mixing timescale (green curve, lower bound of shaded area), the sweep-up timescale (orange curve), and the collision velocities with respect to \vf (purple line, upper bound on shaded area). The black dashed line denotes the assumed \alphat for this model.
\label{fig:constraints}}
\end{figure}

We have implicitly assumed that a stationary distribution is always reached, and thereby neglected some time-dependent effects. A region devoid of small particles only occurs if the population of larger particles has had enough time to remove small grains by sweep-up collisions. Relating the sweep-up timescale \citep[Eq.~A.6]{Birnstiel:2012p18491} to the age of the disk \tdisk, we can derive a condition for the global behavior described above with respect to the turbulence parameter,
\begin{equation}
\alphat > \frac{1}{\left(\Ok\,\tdisk\right)^2} \frac{1}{\epsilon\left|\gamma\right|} \left(\frac{H}{r}\right)^{-2},
\label{eq:sweepup_constraint}
\end{equation}
where $H = \cs/\Ok$ is the pressure scale height and \Ok the Keplerian frequency.
Another criterion for the existence of a `region IV' comes from the assumption that both small and large grains are always in a vertical mixing-settling equilibrium distribution. Without efficient vertical mixing, the inward drifting large grains, located near the mid-plane, would not be able to sweep up the small grain population in the disk atmosphere. Forcing the vertical mixing timescale to be less than the disk lifetime also constrains \alphat, to
\begin{equation}
\alphat > \frac{1}{\tdisk\Ok}.
\label{eq:mixing_constraint}
\end{equation}
Furthermore, \cref{eq:rf_criterion} can be recast as a limit on \alphat,
\begin{equation}
\alphat < \frac{\left|\gamma\right|}{3\epsilon} \left(\frac{\vf}{\Vk}\right)^2,
\label{eq:drift_constraint}
\end{equation}
which ensures that turbulence is too weak to cause fragmentation at a given radius. \Cref{fig:constraints} shows these three conditions for \alphat as function of radius for the parameters of the example simulation (cf., \cref{fig:comparison}). In this case, the sweep-up timescale is not relevant for this simulation. Also, the lower bound on \alphat imposed by \cref{eq:mixing_constraint} is irrelevant throughout most of the disk. The dashed line in \Cref{fig:constraints} corresponds to the adopted \alphat in the simulation. The intersection of \alphat and the upper limit is by definition at \rf and corresponds to the onset of the dip. In cases where \alphat$<10^{-4}$, the regions depleted in small grains can reach down to a few AU, while for $\alphat>10^{-2}$ fragmentation is active throughout most of the disk. \cref{fig:constraints} also shows that the largest possible size of the gap is only weakly dependent on \alphat.

\acknowledgments
We thank M. Benisty and M. Flock for useful discussions.
S. M. A. and T. B. are grateful for support from the NASA Origins of Solar Systems grant NNX12AJ04G, the Smithsonian Institution Pell Grant program, and for generous computing time on the Smithsonian Institution high performance cluster, {\tt hydra}.
Astrochemistry in Leiden is supported by NOVA, KNAW professor prize, and by the European Union A-ERC grant 291141.

\bibliographystyle{apj}

\end{document}